\documentclass[prx,preprint,superscriptaddress,floatfix,aps,showkeys]{revtex4-2}
\usepackage{amsmath}
\usepackage{amsfonts} 
\usepackage{graphicx}
\usepackage{verbatim}
\usepackage[table]{xcolor}
\usepackage[abovefloats]{footmisc}
\usepackage{tabularx}  
\usepackage[export]{adjustbox}
\usepackage{hyperref}

\begin{document}

%\title{2D polaritons for Squeezing and Shrinking Light}

\begin{center}
\textbf{\large 2D polaritons for Squeezing and Shrinking Light}\\[0.5em]

Eduardo B. Barros,$^{1,3*}$, Niclas S. Müller$^{2}$, Stephanie Reich$^{3}$,
Christian Thomsen$^{3}$\\[0.5em]

\textit{$^1$ Department of Physics, Universidade Federal do Cear\'a, Fortaleza,\\ 60455-760 Brazil}\\
\textit{$^2$ Department of Physics, Freie Universit\"at Berlin,\\Berlin, 14195 Berlin, Germany\\}
\textit{$^3$ Department of Physics, Technische Universit\"at Berlin,\\ Berlin, 610101, Germany\\}

\date{\today}
    
\end{center}

%\author{Eduardo B. Barros}\email{ebarros@fisica.ufc.br}
%\affiliation{Department of Physics, Federal University of Ceará, Fortaleza, 60455970, Brazil}
%\author[2]{{Niclas S.} {Müller}}\email{}
%\affiliation{Department of Physics, Freie Univesität Berlin, Berlin, 14195, Germany}
%\author[2]{Stephanie Reich}\email{}
%\affiliation{Department of Physics, Freie Univesität Berlin, Berlin, 14195, Germany}
%\author[3]{{Christian} {Thomsen}}\email{}
%\affiliation{Department of Physics,Technische Universität Berlin, Berlin, 610101, Germany}


Two-dimensional (2D) materials have emerged as an extraordinary platform for polaritonics, given their remarkable ability to confine light to deep sub-wavelength scales. However, as material thicknesses reach the nanometer limit, macroscopic electrodynamic models become inadequate for the description of light-matter coupling in 2D materials, leaving a significant gap in our understanding of 2D polaritons. In this work, we develop a comprehensive quantum microscopic framework that describes 2D polaritons through a general, scale-less Hamiltonian. Starting from the quantum properties of a polarizable 2D surface, we derive collective excitations and  we identify both transverse electric magnetic modes existing outside the light cone. Our results demonstrate that these 2D systems facilitate a powerful synergy between extreme wavelength shrinking and intrinsic quantum squeezing of electric field quadratures. By establishing this theoretical roadmap, our work provides the essential architecture for controlling non-classical states of light at the nanoscale, opening new avenues for high-precision quantum sensing and the development of next-generation optoelectronic devices

\newpage

\maketitle

\section{Introduction}

%Introduce polaritons in 2D materials and discuss importance and relevance and the importance of modelling and understanding

% Opening: What are the big questions 

%Polaritons in bulk materials are hybrid light-matter quasi-particles  that are exploited to shrink and confine light to sub-wavelength scales with applications in opto-electronics, high-resolution light sensing, and analytics.\cite{Wu2022,Epstein2020,Rivera2019} In addition, the quantum properties of polaritons, such as non-Poissonian photon distributions and squeezed quadratures,\cite{Walls1983,Artoni1991} have been studied since the seminal work by Hopfield in the late 50s,\cite{Hopfield1958,Bazov2025}, but harvesting them for applications remains challenging. 

%Polaritons in two-dimensional (2D) materials offer unique opportunities for accessing the electric fields  and quantum properties of polaritons, because their evanescent fields extend well into the surrounding environment where they can be probed experimentally.\cite{Rivera2016,Koenderink2015}  
%     

Polaritons are formed by the coupling between collective matter excitations, such as plasmons, phonons, or excitons and the electromagnetic field. In three-dimensional (3D) materials polaritons are separated into bulk and surface polaritons.\cite{Mills1974} Bulk polaritons propagate through the material and determine its electromagnetic response. Surface polaritons are bound to material interfaces and progagate along them. They are associated with evanescent electric fields that decay rapidly away from the surface.\cite{Mills1974,Pitarke2007} 

A problem arises when trying to describe polaritons in ultra-thin materials slabs ($<10$ nm): The thickness of the slab is an order of magnitude smaller than the wavelength of light ($\lambda>300$\,nm). Under this condition, the macroscopic Maxwell equations become ill suited for describing the electromagnetic response, leading to apparently vanishing polaritonic effects for nanometer slab thicknesses.\cite{Hu2017,Kusch2021}  Recent descriptions based on microscopic Maxwell equations addressed this problem using surface conductivities and susceptibilities to describe ultra-thin materials.\cite{GoncalvesBook,Rivera2019, Khurgin2015, Li2024}.  

Recent works have highlighted quantum nanophotonics for the realization of exquisite phenomena such as quantum nonlocality, ultrastrong light–matter coupling, forbidden transitions, polaritonic quantization, and topological effects. Despite this, a deep understanding of 2D polaritons within a quantum mechanical framework is currently lacking.  In this work, we fill this gap by providing a general quantum description of 2D polaritons and their associated electric fields. Our results open a route for targeting and controlling the quantum nature of 2D polaritons.  

%This approach were applied to plasmon polaritons in graphene \cite{GoncalvesBook}  and phonon polaritons in few-to-monolayer boron nitride and is  

The quantum properties of bulk and surface polaritons have been described previously,\cite{Hopfield1958,Quattropani1986,Artoni1991,Lindel2021,Alpeggiani2014} but this framework has not been extended to 2D polaritons. Clearly, the top-down strategy --starting from a quantum version of macroscopic Maxwell equations and  applying boundary conditions for slabs of vanishing tickness-- 
is just as inadequate from a quantum perspective as it was for the classical one. We propose adopt the alternative strategy inspired by the microscopic classical approach: Start from the quantum description of a polarizable 2D surface and add light-matter interaction to arrive at 2D polaritonic quasiparticles. We find that 2D systems sustain transverse electric (TE) and transverse magnetic (TM) polaritons with electric fields confined to the region near the 2D plane. The 2D polaritons 
shrink the wavelength of the electromagnetic field by XXX compared to a vacuum photon of the same frequency. They can also be used to produce squeezed light for quantum optical applications such as high precision sensing.

\section{Model \label{sec:model}}

We developed a quantum mechanical description of 2D material excitations and their interaction with light. We start from a 2D polarizable infinitesimal plane with a surface charge density $\sigma_s$. The bare in-plane excitations are assumed to oscillate with frequency $\omega_T$. Two types of collective in-plane dipole excitations occur in these 2D systems, longitudinal and transverse  modes, where the charge displacement is either parallel or perpendicular to the propagation direction, Fig.~\ref{fig:dispersion}\textbf{a}. For the transverse mode, the charge density modulation does not generate an electric field, leading to a non-dispersive mode with frequency $\omega_T$, see $\S$\ref{sec:methods}. 
In the case of longitudinal excitations, however, the charge density modulation generates a non-zero electric field that evanesces away from the plane. This non-retarded field arises from the scalar Coulomb potential and renormalizes the longitudinal mode frequency, see $\S$\ref{sec:methods},
\begin{equation}\label{eq:omegaL}
\omega_L(k)=\sqrt{\omega_T^2+g^2_\mathrm{2D}k\lambda_T}, 
\end{equation}
where $k$ is the mode wavevector and $\lambda_T$ ...
The in-plane Coulomb interaction is parameterized by the coupling parameter
\begin{equation}
g_\mathrm{2D}=\sqrt{\frac{\sigma_s e^2}{4 m\epsilon_0 \lambda_T}}   
\end{equation}
that depends on the surface charge density and the effective mass of the excitations.

The collective excitations couple to the free-space photons leading to the formation of 2D polaritons. We describe this coupling in terms of a Hopfield-like \cite{Hopfield1958} model for the light-matter interaction, which is extended to include the free photons ($\S\ref{sec:methods}$). The photons are separated into $s$- and $p$-polarized. The electric fields of $s$ polarized photons are parallel to the 2D surface and perpendicular to the in-plane component of the wavevector; they couple only to transverse matter excitations. The electric field of $p$ polarized photons, in contrast, has components along the in-plane wavevector and in the out-of-plane direction; such photons couple the the longitudinal in-plane modes. Interestingly, the light-matter coupling strength for both types of excitations is governed by the parameter $g_\mathrm{2D}$ introduced for the dispersion of longitudinal matter modes above, highlighting the close connection of light-matter and matter-matter interactions.\cite{Barros2024} 
The obtained Hamiltonian ($\S$ \ref{sec:methods}) is quite general and intrinsically scale-less. It can be used to describe any type of 2D polariton. The time scale of the system is defined by the frequency of the transverse in-plane excitations $\omega_T$ while the distances are defined by the speed of light in the surrounding medium (considered here to be vacuum). 
The Hamiltonian is expressed in terms of creation and annihilation operators for the bare matter and electromagnetic excitations. It is diagonalized by a Bogoliubov transformation, see $\S$4.

\section{Results}
\begin{table}
    \centering
    \begin{tabular}{ccccc}
       \hline\hline \begin{tabular}{c}Mode\\type\end{tabular}  & \begin{tabular}{c}Matter\\excitation\end{tabular}&\begin{tabular}{c}Photon \\excitations\end{tabular} & \begin{tabular}{c}Electric \\fields\end{tabular} & \begin{tabular}{c}3D \\counterpart \end{tabular}   \\ \hline
       TM   & longitudinal &   p-polarized & \begin{tabular}{c} longitudinal \\(in-plane)\\transverse \\(out-of-plane)\end{tabular} & \begin{tabular}{c} surface \\ polaritons \end{tabular} \\ \hline
       TE   & transverse &  s-polarized & \begin{tabular}{c} transverse \\(in-plane) \end{tabular} &\begin{tabular}{c} bulk \\polaritons \end{tabular} \\\hline\hline
    \end{tabular} 
    \caption{Summary of the properties of the TE and TM 2D polaritons}
    \label{tab:my_label}
\end{table}

\begin{figure*}
    \centering
  %  \includegraphics[width=4.2cm]{2Dpolariton_model.png}
 %   \begin{subfigure}
    \includegraphics[width=7cm]{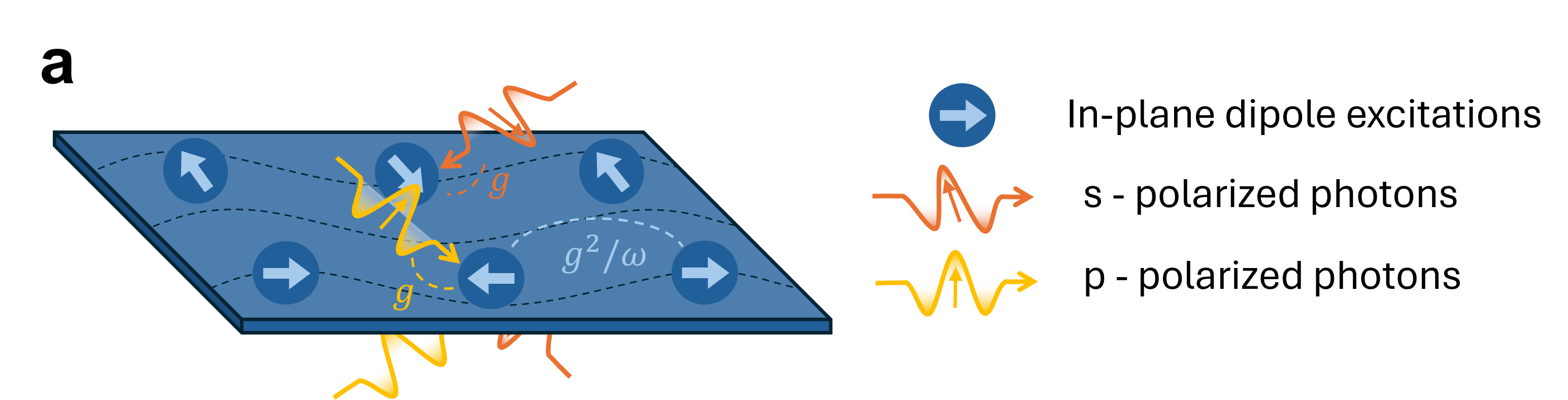}     
  %  \end{subfigure}
  %  \begin{subfigure}
        
    \includegraphics[width=4cm]{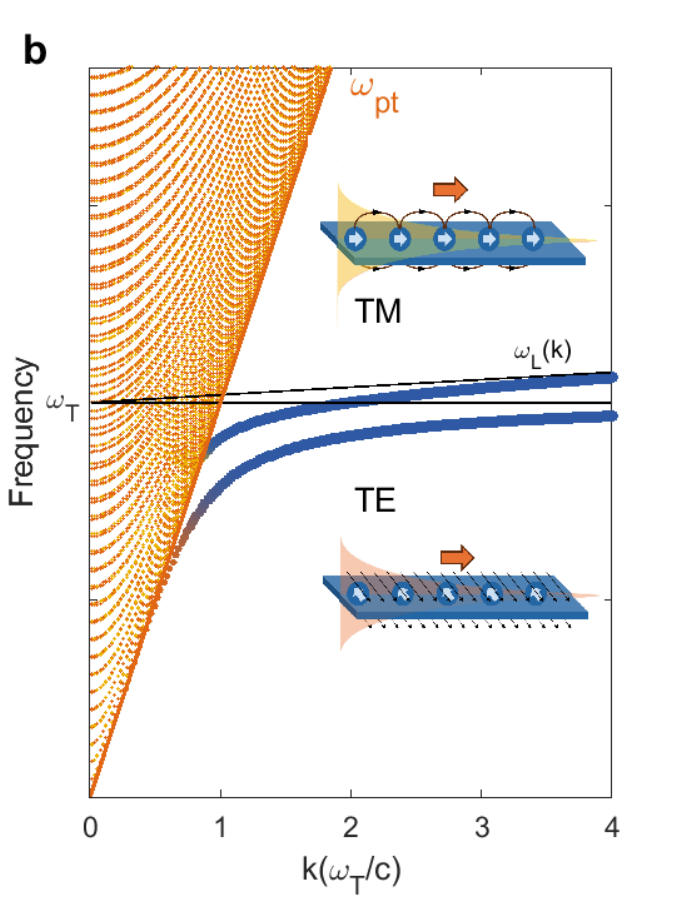}
     \includegraphics[width=4cm]{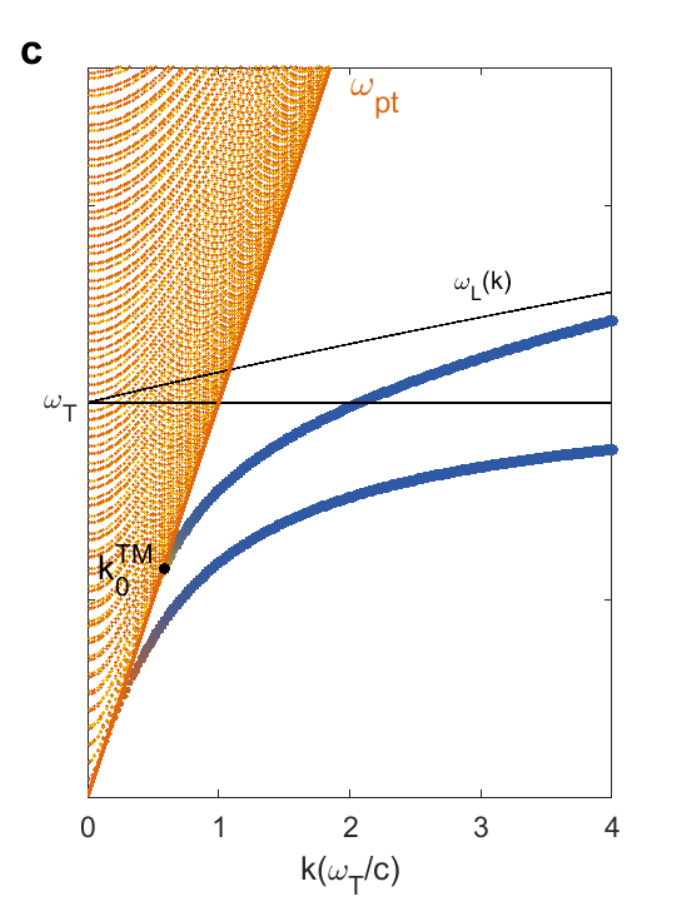}
    \includegraphics[width=4cm]{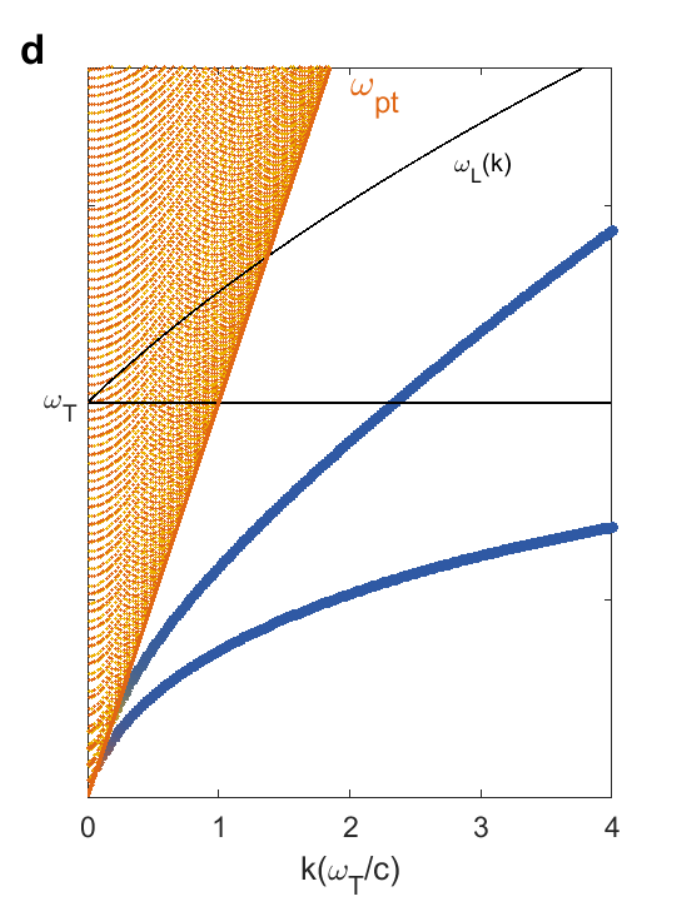}
 %\end{subfigure}
    
 %  \includegraphics[width=9cm]{Fig_Emap_hop.pdf}
    \caption{\textbf{a} In-plane localized surface excitations couple to each other and to $s$- and $p$-polarized photons. Polariton dispersion for $g_\mathrm{2D}/\omega_T=$ \textbf{b}~0.2, \textbf{c}~0.4 and \textbf{d}~0.8. Symbol colors represent the contribution of matter (blue) excitations, $s$ (orange) and  $p$ (yellow) polarized photons to the eigenstates of the polaritonic system. Black lines are the longitudinal ($\omega_L$) and transverse ($\omega_T$) dispersions without coupling to the free photons.   Sketches in \textbf{b} show TE modes and TM modes.  $k_0^\mathrm{TM}$ in \textbf{c} represents the smallest wavevector of the TM polariton; it decreases with increasing coupling. }
    \label{fig:dispersion}
\end{figure*}

 The quantum mechanical light-matter Hamiltonian for 2D materials predicts a series of real frequency bands $\omega_{pol}(k)$ corresponding to the excitations of the system. Most of these modes are nearly free photons with a nearly parabolic dispersion, yellow and orange  bands in Fig.~\ref{fig:dispersion}. In addition, two polaritonic states appear as solutions outside of the light-cone in reciprocal space ($k>\omega_{pt}/c)$, blue bands in Fig.~\ref{fig:dispersion}. The lower frequency state is a transverse electric (TE) mode that emerges from the  interaction of  $s$-polarized photons and  transverse matter excitations in the 2D monolayer. This mode is reminiscent of the bulk transverse polariton in 3D materials and correlates well with the 2D TE polariton discussed by Khurgin.\cite{Khurgin2015} Similarly to the bulk polaritons, the dispersion of the TE polariton smoothly changes from a photon-like dispersion near $k=0$ to the matter-like dispersion ($\omega=\omega_T$) for large $k$. Characteristic properties of the 2D polaritons are summarized in Table~\ref{tab:my_label}.    

The higher frequency mode in the non-retarded region is the transverse magnetic polariton (TM), Fig.~\ref{fig:dispersion}. It is reminiscent of a surface polariton at the interfaces between 3D systems. This TM polariton is formed by the coupling of the in-plane longitudinal matter excitations to $p$-polarized photons. Its frequency  converges towards the bare longitudinal matter mode large $k$. In this limit, it can be described by considering only the Coulomb interaction between the localized in-plane dipole excitations, effectively disregarding their interaction with light. For small $k$, however, the TM polariton frequency gets dressed by the free photons causing a frequency redshift. The characteristic wavevector $k_0^\mathrm{TM}$ at which the TM polariton frequency crosses the light cone shifts down compared to the uncoupled case, another effect of the light-matter coupling. 

The frequency range of the TM mode of 2D materials is larger than the frequency range of the 3D surface polaritons, creating an important distinction between bulk and 2D materials. In 3D systems the minimum frequency of the surface polariton, $\omega_0^\mathrm{TM}=ck_0^\mathrm{TM}$, always coincides with the frequency of the bare transverse mode. Surface polaritons exist only within the gap created by the splitting of the collective longitudinal and transverse matter mode called the Reststrahlen band of the crystal. In  2D materials, however, $\omega_0^{TM}$ becomes much smaller than $\omega_T$ for sufficiently large light-matter coupling $g_\mathrm{2D}$. For extreme magnitudes of the reduced light-matter coupling, the onset of the longitudinal polartion can shift all the way down to zero, Fig.~\ref{fig:dispersion}\textbf{c}, due to the interaction with short-wavelength (high energy) free-photons.  

\begin{figure*}
    \centering
    \includegraphics[width=4cm]{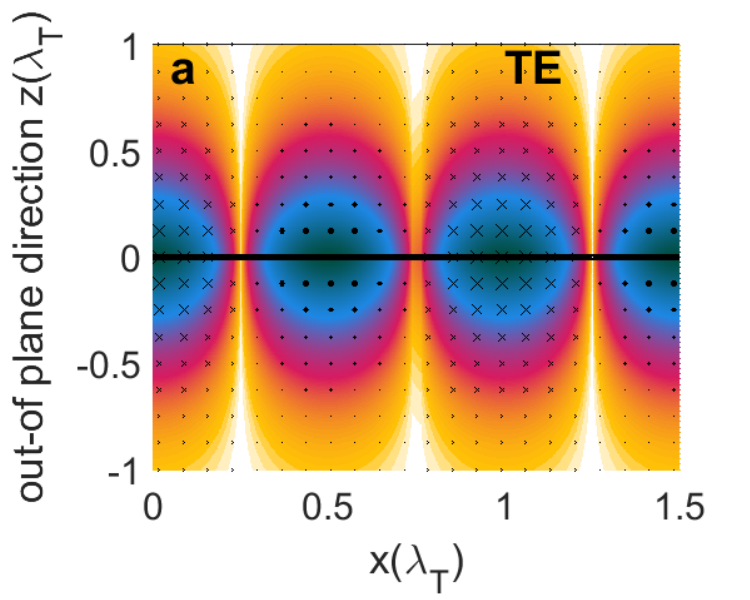}
        \includegraphics[width=4cm]{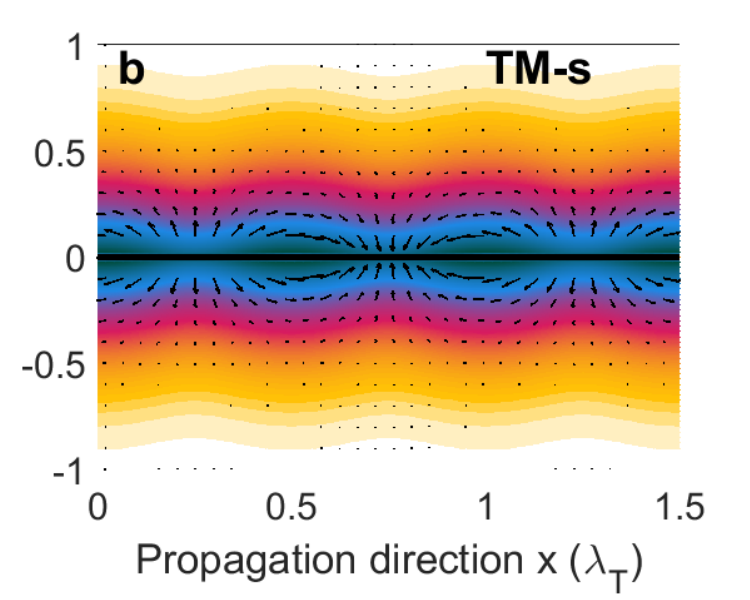}
            \includegraphics[width=4cm]{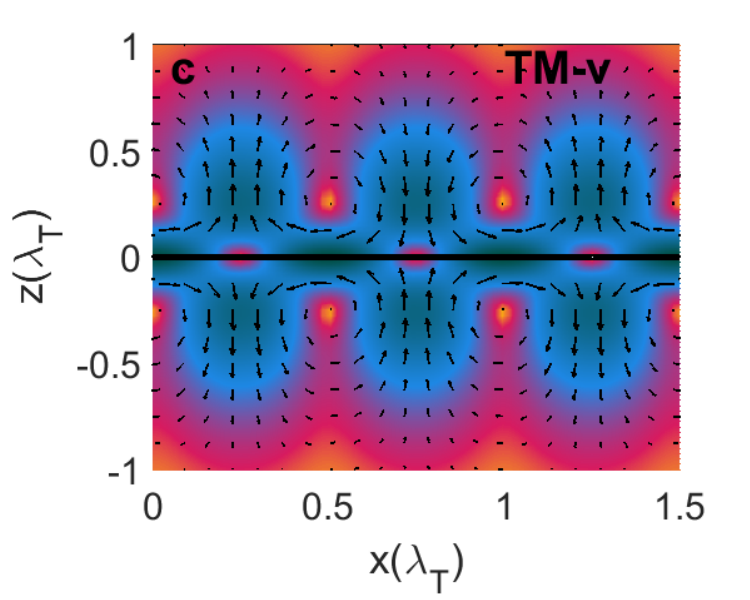}
          \includegraphics[width=6cm]{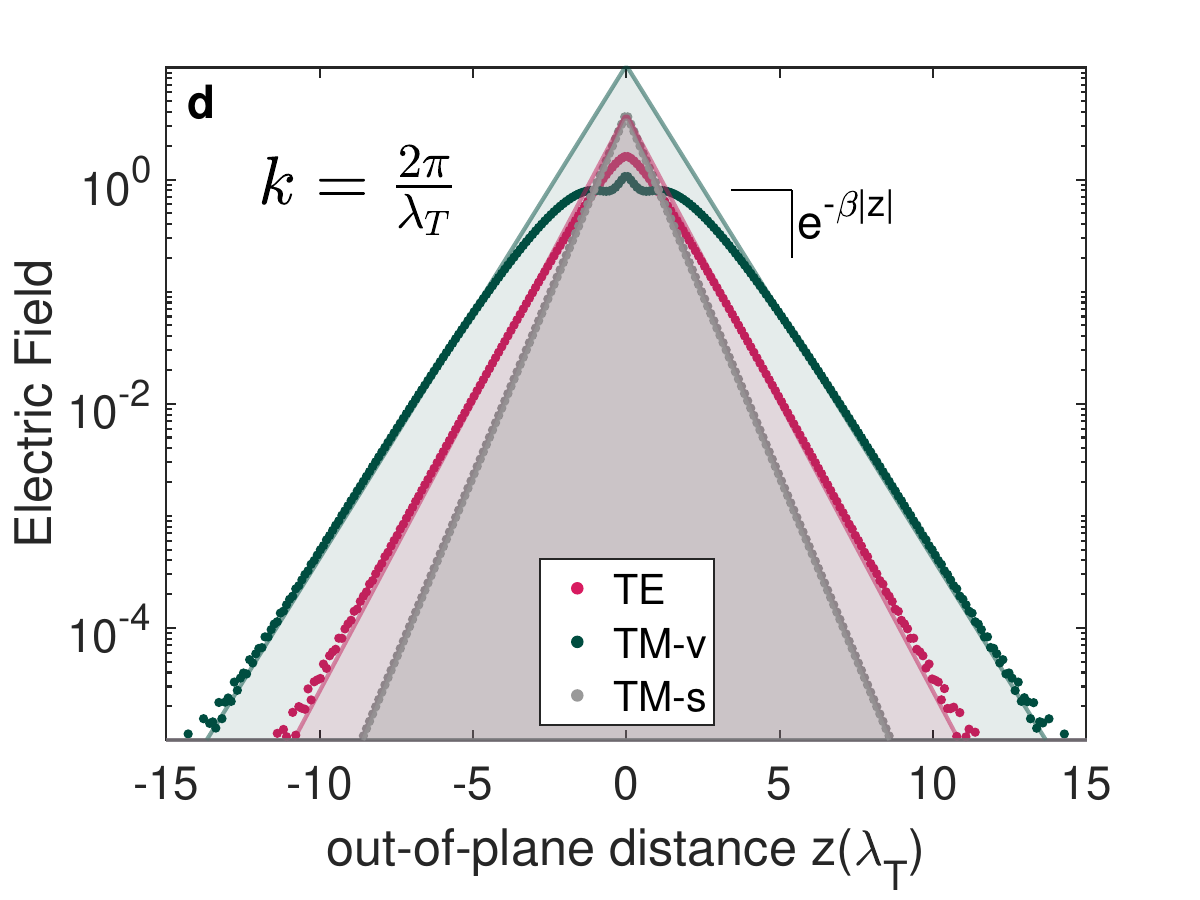}
    \includegraphics[width=6cm]{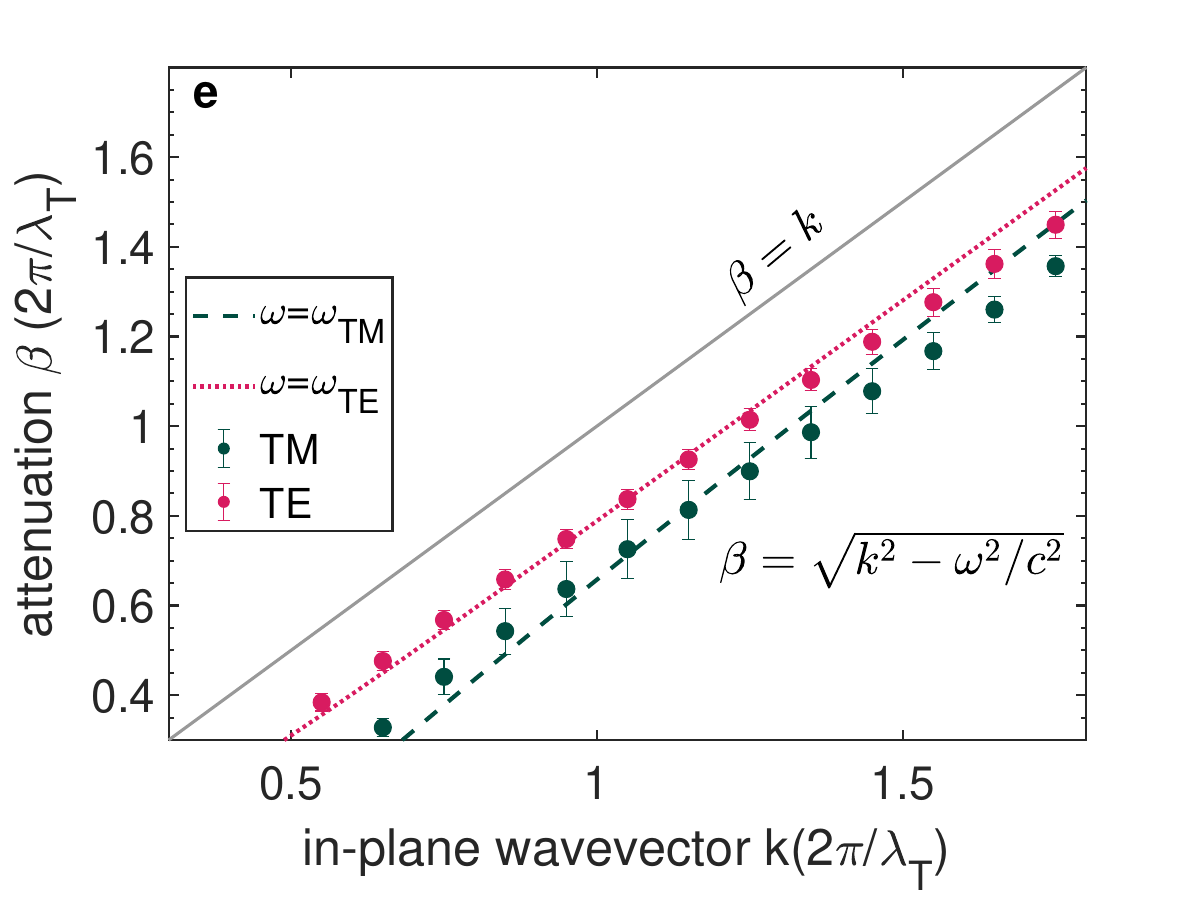}    
        \caption{Expectation value for the \textbf{a} In-plane Transverse ($E_T$), \textbf{b} in-plane longitudinal ($E_L$) and \textbf{c} out-of-plane ($E_{out}$) components (evaluated at $z=0.025\lambda_0$). The dark horizontal line represents the 2D material. \textbf{d} Spatial distribution of the expected electric field for the marked points in $\textbf{a}-\textbf{c}.g/\omega_T=0.4$. \textbf{e} Dependence of the attenuation $\beta$ on the in-plane wavevector $k$. For small $k$ the obtained behavior befits the expected classical results. Large $k$ deviations are mainly due to numerical inaccuracy.}
    \label{fig:2}
\end{figure*}

Quantum effects lead to a non-trivial modulation of the localized electric fields for the TM polariton. To depict this, we consider the spatial distribution of the electric fields for coherent TE and TM polaritons. The quantum approach gives expressions for the electric field operators $\hat E(\mathbf r,t)$,(see $\S$.\ref{sec:methods}) that we use to describe their quantum properties. The expectation values of these operator is calculated for a chosen quantum state. The choice of coherent states is justified as these are the closest to classical states and yield real expectation values for electric fields. 
%The spatial distribution of the localized electric fields associated to the TE and TM polaritons differ greatly. To depict this, we calculate the expectation value of the electric field operator $\langle\mathbf E^+\rangle_{\tau,\mathbf k} $ on a coherent polariton state 
%$|\tilde\alpha_{\tau,\mathbf k}\rangle$, ($\tau=TM,~TE$). % to calculate the expectation %.  =\langle\tilde\alpha_{\tau,\mathbf k}| \mathbf E^+|\tilde\alpha_{\tau,\mathbf k}\rangle$. 
%We choose coherent polarithe coherent states $|\tilde\alpha_{\tau,\mathbf k}\rangle$ are defined as the eigenstates of the polariton annihilation operators $\psi_{\tau,\mathbf k}|\tilde\alpha_{\tau,\mathbf k}\rangle=\tilde\alpha_{\tau,\mathbf k}|\tilde\alpha_{\tau,\mathbf k}\rangle$, and  with $\tilde\alpha_{\tau,\mathbf k}$ complex. 
%Expressions for the electric field operator $\mathbf E(\mathbf r,t)=-\dot{\mathbf A}(\mathbf r,t)-\nabla\phi(\mathbf r,t)$, with both its vector and scalar potential components, are shown in the theory session. % in terms of polaritonic creation and annihilation operators. 
The electric field produced by a coherent TE polariton (Fig. \ref{fig:2}\textbf{a}) originates exclusively from the interaction OF WHAT to the vector potential. This creates components that are simultaneously perpendicular to the in-plane propagation direction and the out-of-plane $z$ direction. The field oscillates sinusoidally in the propagation direction with a wavelength $\lambda=2\pi/k$ while decreasing along $z$ as $e^{-\beta |z|}$, pink curve in Fig.~\ref{fig:2}\textbf{d}, with the attenuation $\beta_{TE}\sim \sqrt{k^2-\omega^2_{TE}/c^2}$ (pink spheres in \ref{fig:2}\textbf{e}). 1-2 sentence of intepretation/comparison to expectations.

\begin{figure*}[t]
    \centering
    \includegraphics[width=5cm]{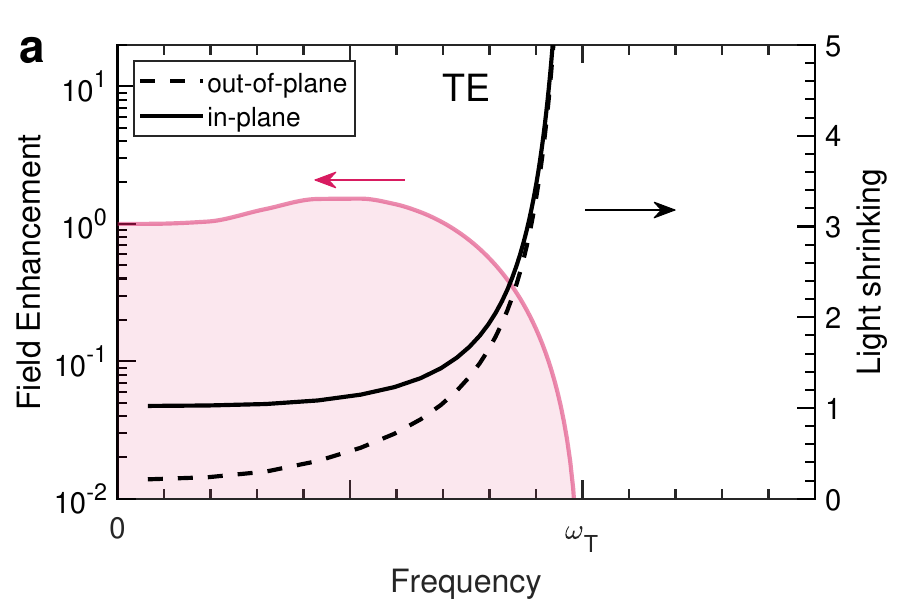}
            \includegraphics[width=5cm]{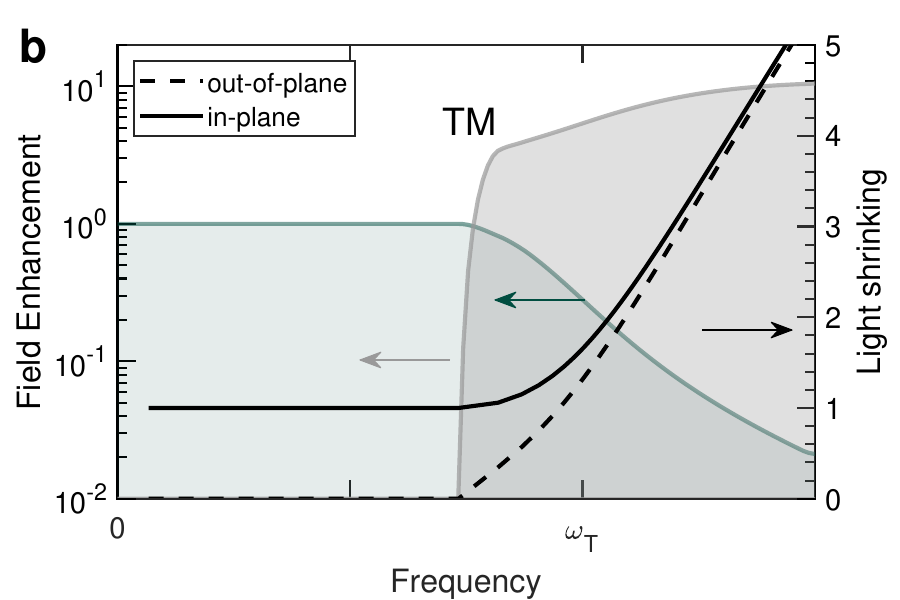}
    \includegraphics[width=5cm]{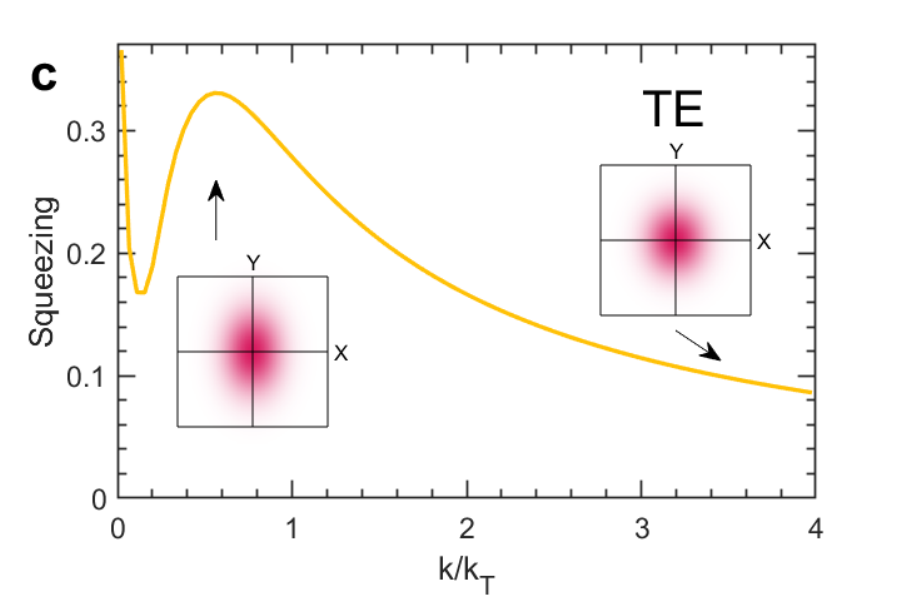}
     \includegraphics[width=5cm]{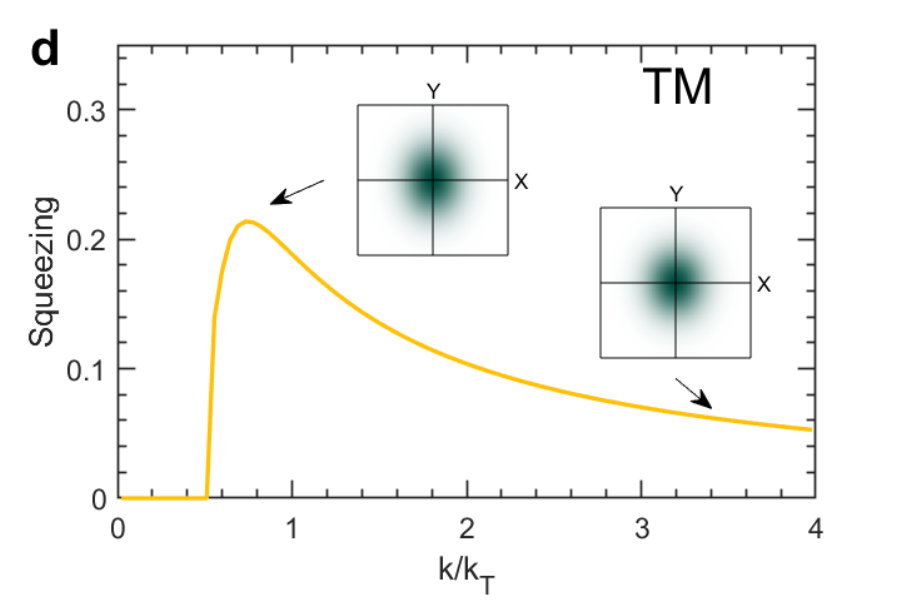}
        \includegraphics[width=5cm]{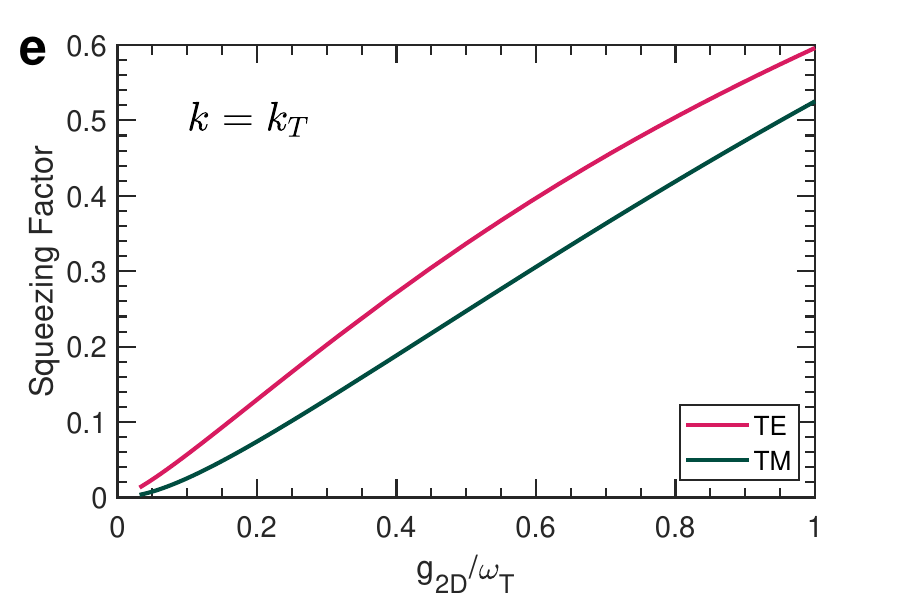}  
               \includegraphics[width=5cm]{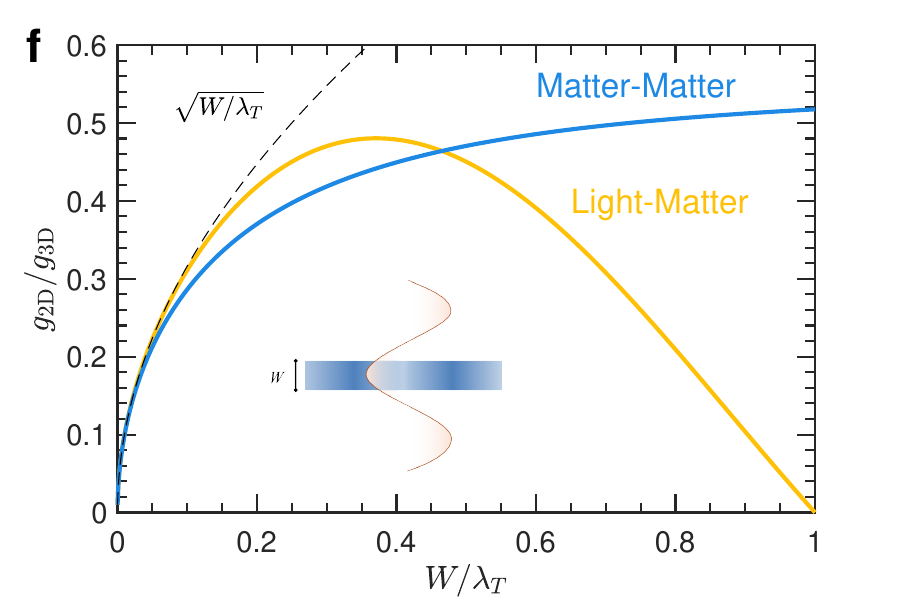}  
     
    \caption{Polariton electric field magnitude (pink/green areas) , in-plane (dark lime) and out-of-plane (dashed line) light shrinking  for the \textbf{a} TE  and  \textbf{b} TM 2D polaritons. The gray area in \textbf{b} represents contributions from the scalar potential to the TM polariton. Electric fields are evaluated at $z=0$. 
    Squeezing magnitude, see \ref{sec:methods} for details, of the \textbf{c} TE and \textit{d} TM polaritons. The sketches depict the Wigner distribution (assumed Gaussian) of the polaritonic ground state at specific values of $k$. \textbf{e} Nearlu linear dependence of the squeezing factor on $g_\mathrm{2D}/\omega_T$ for both TE and TM polaritons evaluated at $k=k_T$. \textbf{f} Effective light-matter (yellow) and matter-matter coupling parameters as a function of the material width    }   
    \label{fig:ShrinkSqueeze}
\end{figure*}

The electric field produced by the coherent TM polariton in 2D materials has contributions from the vector and scalar potentials (Figs.~\ref{fig:2}\textbf{b} and \textbf{c}). Each of these contributions is composed of longitudinal and out-of-plane components. The contribution by the scalar potential  to the electric field gives rise to the longitudinal mode renormalization discussed in $\S$~\ref{sec:model}. It shows a sinusoidal modulation in the propagation direction and an exponential decay along $z$, similar to the electric field produced by the TE polariton. The field produced by the vector potential, in contrast, shows a non-trivial spatial modulation in the out-of-plane direction. This results from different dependencies of the longitudinal $E_L$ and out-of-plane component $E_z$ on the $z$ leading to a structured electric field with well defined vortices around points of zero electric field (see Fig.~\ref{fig:2}\textbf{c}). This modulation is particularly relevant in the close vicinity of the 2D material $z<<\lambda_T$ and large coupling strengths $g_\mathrm{2D}/\omega_T>0.1$. For large $z$ and weak coupling, the electric field becomes similar to the scalar component, thus recovering the classically expected result. The attenuation rate for the electric field produced by the TM polariton (including the scalar and vector components) correspond to the classical value $\beta_{TM}=\sqrt{k^2-\omega_\mathrm{TM}^2/c^2}$ for large $z$ (green spheres in \ref{fig:2}\textbf{e}). The vector potential contribution is then essential for obtaining the expected attenuation rate, highlighting the important role played by the light-matter interaction on defining the polaritonic electric fields for both TE and TM polaritons.

The  electric fields created by the polaritons are highly localized in the out-of-plane and  the in-plane propagation directions. We quantify this effect by the shrinking factor $s_k$ that compares the wavelength of the field distribution  to the wavelength of a photon with the same frequency in free space.\cite{Rivera2016,Iranzo2018,Epstein2020} For the TE polariton the shrinking factors are close to one far away from the bare transverse frequency, Fig.~\ref{fig:ShrinkSqueeze}\textbf{a}, and 
increase sharply close to $\omega_T$. For the TM polariton the shrinking starts to rise at $\omega_\mathrm T$ and keeps increasing for higher energies (Fig.~\ref{fig:ShrinkSqueeze}\textbf{b)}.  
To exploit light shrinking experimentally and produce a highly localized electric response, the polariton must also create a finite electric field. For the TE mode, the maximum electric field intensity (pink area in Fig.~\ref{fig:ShrinkSqueeze}\textbf{a}) vanishes supra-exponentially as the frequency approaches $\omega_T$. Large shrinking factors ($s_k>10$) become inaccessible for the TE polariton. For the TM polariton, however, the scalar field contribution dominates over the vector potential in the frequency range with large shrinking factors. For large wavectors, the scalar potential becomes sufficient to describe the shrinking power of 2D polaritons.\cite{GoncalvesBook,Hu2017,Dai2014} The photonic contribution to the TM electric field is only relevant near the light cone, where the confinement is small ($s_k \sim 1$).   

The electric fields created by 2D polaritons are intrinsically squeezed and can be used for the production of non-classical light. To show this, we  separate the electric field operator $\boldsymbol{E}$ associated to each polariton state ($\tau=$TE, TM) into two quadrature operators, $\hat E_\tau(\mathbf r,t)= \sum_{\mathbf k} |E_{\tau}(\mathbf k)|[\hat{\mathcal C}_{\tau,\mathbf k} \cos(\omega_{\tau,\mathbf k} t)+ \hat{\mathcal S}_{\tau,\mathbf k} \sin(\omega_{\tau,\mathbf k} t)]/\sqrt{2}$, each oscillating either as the cosine ( $\hat{\mathcal C}$) or the sine ( $\hat{\mathcal S}$) of $\omega_{\tau,\mathbf k} t$. We evaluate the intrinsic polariton squeezing as the quantum fluctuations of the two electric field quadratures in the bare photon ground state.\cite{Artoni1991}  This justified by the fact that in most experimental setups, the polaritons will be excited by a source of coherent free photons, such as an incident laser beam. 2D polaritons are intrinsically squeezed for $k\approx k_T$, corresponding to the wavevectors where the hybrid light-matter character of the polariton is most pronounced. As $k$ increases, the dressing of the polaritons by the photons weakens, and the squeezing starts to decrease, become negligible for $k>>k_T$. The magnitude of the squeezing increases with increasing light-matter coupling strength. For instance, for $k=k_T$ the squeezing factor scales approximately with $g_\mathrm{2D}$ to reach values of more than 0.5 as $g_\mathrm{2D}/\omega_T$ approaches unity, Fig.~\ref{fig:ShrinkSqueeze}\textbf{e},

One challenge for observing quantum properties with 2D polaritons lies in realizing large enough coupling factors $g_\mathrm{2D}$. Squeezing factors of $\sim0.2$ (corresponding to a difference of about 20\% in the quantum fluctuations of the two quadratures) are obtained for $g_\mathrm{2D}\approx 0.2 \omega_T$ for the TE (TM) mode, Fig. \ref{fig:ShrinkSqueeze}\textbf{e}.  Our proposed model helps in designing strategies to achieve such coupling constants. For small thicknesses $W<<\lambda_T$ the 2D coupling constant is approximately $g_\mathrm{2D}=g_\mathrm{3D}\sqrt{W/\lambda_T}$, where $g_\mathrm{3D}$ is the light-matter coupling constant of the corresponding bulk material. This allows us to estimate the coupling constant for phonon polaritons in monolayer boron nitride to be roughly 100 times smaller than that of bulk hBN. One strategy for increasing $g_\mathrm{2D}$ is to change the energy range of the excitation. By moving to exciton-polaritons, the light-matter coupling constant for an MoS$_2$ monolayer  is 30 times smaller than for bulk MoS$_2$. Higher frequency excitations, such as UV excitons or plasmon-polaritons may realize even stronger coupling. Recent experiments on plasmonic supercrystals showed that the light-matter coupling in few-layer of hexagonally packed gold nanoparticles is identical to the bulk nanoparticle supercrystal.\cite{mueller2020deep} 

Another possible strategy is to inc
rease the thickness of the 2D material, which increases the total charge density and results in stronger light-matter coupling. Charge oscillations in a slab with a finite thickness $W$ yield an effective coupling to free photons, see Supplementary Information
\begin{equation}
g^{LM}_\mathbf{2D}=g_\mathrm{3D}\frac{\sin(\pi W/\lambda_T)}{\pi\sqrt{W/\lambda_T}}.
\end{equation} 
As  $W$ increases, the ratio between the 2D and 3D coupling parameters has an upper limit $g^{LM}_\mathbf{2D}/g_\mathrm{3D}=\sqrt 2/\pi\sim 0.35$ that is reached for $W=\lambda/2$. This limit assumes a homogeneous density of oscillating charges  across the width of the slab, an approximation that fails for thicker slabs, where different charge oscillation profiles are expected. The in-plane Coulomb interaction is also modified by the thickness of the 2D material, such that the renormalization of the longitudinal mode is parameterized by the effective coupling constant $g^{MM}_\mathrm{2D}=g_\mathrm{3D}\sqrt{\frac{1}{\pi}\left(1-\frac{\sinh{(\pi W/\lambda_T})}{\pi W/\lambda_T}\right)}$. Figure \ref{fig:ShrinkSqueeze}\textbf{f} shows the evolution of the effective light-matter and matter-matter coupling parameters as a function of the slab width $W$. For $W<0.02 \lambda_T$, where the approximation $g^{LM}_\mathbf{2D}=g^{MM}_\mathbf{2D}\approx g_\mathbf{3D}\sqrt{W/\lambda_T}$ is valid, should be seen as the validity range of the present model. Values of $g_\mathrm{2D}$ only 10 times smaller than the bulk coupling parameter are readily achievable by increasing the number of layers, indicating that materials with bulk light-matter coupling in the deep-strong coupling regime ($g_\mathrm{3D}/\omega_T>1$) are excellent candidates for the observation of quantum polaritonic properties.

The final strategy for reaching the quantum regime in 2D polaritonics are materials where the frequency of the  transverse excitations approaches zero. This is the case for the plasmons of graphene and other 2D conductive materials, because there are no restoring forces for the movement of their free electrons. The plasmon-polariton dispersion of graphene is described by our model when taking the limit $\omega_T\rightarrow 0$. In this limit the wavelength $\lambda_T$ is ill-defined, but the  coupling parameters can be evaluated for infinitesimal $\omega_T$. Direct comparison to the classical graphene dispersion leads to $g^2_{graph}\lambda_T\rightarrow =2\alpha \varepsilon_F$, where $\alpha$ is the fine structure constant and $\varepsilon_F$ is the Fermi energy \cite{GoncalvesBook}. As the quantum effects of light-matter coupling increase with the ratio $g_\mathrm{2D}/\omega_T$, they will become relevant in such systems. Possible quantum effects and its applications should be studied in greater detail for plasmon-polaritons in 2D materials.

In conclusion, we developed a foundational quantum microscopic framework for the description of 2D polaritons, effectively bridging the critical gap between classical electrodynamics and the intrinsic quantum nature of light-matter interactions at the nanoscale. The quantum description of the polaritonic electric fields localized around the 2D material informs on the necessary experimental framework for quantum polaritonics, working as a forefront for future scientific advances.

\section{Theory}

%A free-standing layer can be described by extending the Hopfield model to consider two degenerate in-plane bosonic excitations in a planar structure at $z=0$ which are allowed to coupled to the free photons in 3D space. To discretize the description, we impose periodic boundary conditions along the $z$ direction.

\label{sec:methods}

We consider here in-plane bosonic surface charge density excitations. The bosonic excitations have an in-plane wave vector $\mathbf{k}=k_x \hat{\mathbf{x}} + k_y \hat{\mathbf{y}}$, while photons can also have an out-of-plane component $\mathbf{q}_z$. The periodic boundary condition in $z$ limits the values of $\mathbf q_z=q \hat{\mathbf{z}}$ to $q=j(2\pi/L_z) $, with $-N_z/2+1<j\leq N_z/2$, where $L_z=N_z \lambda_T$. The free-standing layer is obtained in the limit of $L_z\rightarrow \infty$.

The full hamiltonian is 
\begin{equation}
\mathcal H=\mathcal H_{L} + \mathcal H_M +\mathcal H_{LM} 
\end{equation}
%i
Where $\mathcal H_L$ is the Hamiltonian for free photons, $\mathcal H_M$ is the hamiltonian describing the in-plane excitations and $\mathcal H_{LM}$ is the light-matter interaction. The matter part has two contributions, the free in-plane excitations $\mathcal H_{dipole}$ and the Coulomb 
\begin{equation}
H_{coul}= -\frac{1}{A}\int d^2\rho\hat{\mathbf p}(\boldsymbol\rho)\cdot \hat{\mathbf E}_{sc}(\boldsymbol\rho),    \end{equation}which is the energy change of the surface charge distribution with dipole moment $\hat{\mathbf p}(\boldsymbol{\rho})$  due to the electric field it produces $\mathbf E_{sc}(\boldsymbol\rho,z=0)$. Here $A$ is the area of the material, and $sc$ stands for $scalar$, as this term arises from the scalar potential $\phi(\boldsymbol r)$ part of the electromagnetic potential. This non-local Coulomb interaction affects only the longitudinal mode $\sigma=L$ and leads to a $k$ dependent frequency $\omega_L(k)$. The frequencies $\omega_T(k)$ and $\omega_L(k)$ are shown in Fig.~\ref{fig:2} as thin black lines.      

The full Hamiltonian after the frequency renormalization is  
\begin{equation}
\begin{split}\label{Eq_HFS}
\mathcal{H} = & 
 \sum_{\mathbf{k},\mathbf q_z \lambda}  \hbar \omega_{\mathrm{pt}}(\mathbf{k},q)a_{\mathbf{k},q\lambda}^\dagger  a_{\mathbf{k},q \lambda} + \sum_{\mathbf{k}, \sigma} \hbar \omega_\sigma(\mathbf k) b_{\mathbf{k}, \sigma}^\dagger  b_{\mathbf{k}, \sigma}
\\  + \sum_{\mathbf{k},q,\lambda, \sigma}& i \hbar g_\mathrm{2D} \sqrt{\frac{\lambda_T}{L_z}} \xi_{\mathbf{k},q}^{\lambda,\sigma} (a_{\mathbf{k}, q \lambda}^\dagger + a_{-\mathbf{k},-q\lambda})(b_{-\mathbf{k}, \sigma}^\dagger - b_{\mathbf{k}, \sigma})
\\  + \sum_{\mathbf{k},q,q^\prime, \lambda, \lambda'} & \hbar \frac{g^2_\mathrm{2D}}{\omega_\sigma }\frac{\lambda_T}{L_z} \Xi^{\lambda,\lambda'}_{\mathbf{k}, q, q^\prime} (a_{\mathbf{k},q \lambda}^\dagger + a_{-\mathbf{k},-q \lambda})(a_{-\mathbf{k},- q^\prime \lambda'}^\dagger + a_{\mathbf{k},q^\prime \lambda'}),
\end{split}
\end{equation}
in terms of the creation and annihilation operators for photons ($\hat a^\dag_{\mathbf{k},q\lambda},~\hat a_{\mathbf{k},q\lambda}$) and in-plane matter excitations ($b^\dag_{\mathbf{k},\sigma},~b_{\mathbf{k},\sigma}$). Here $\sigma=L,T$ is for the longitudinal and transverse excitations in-plane excitations and $\lambda=s,p$ for the two possible photon polarizations.  The light matter and Coulomb interactions depend on the same parameter
\begin{equation}
g_\mathrm{2D}=\sqrt{\frac{\sigma_s e^2}{4 m\lambda_T\epsilon_0 }},  
\end{equation}
depending mainly on the surface density $\sigma_s$ of dipole charges in the layer and their effective masses $m$.  

%$=\mathrm q\sqrt{\frac{\hbar}{2m\omega_T}}\sum_{\sigma,\mathbf k}(\hat b_{\mathbf k,\sigma}e^{i\mathbf k\cdot \boldsymbol \rho}+\hat b^\dag_{\mathbf k,\sigma}e^{-i\mathbf k\cdot \boldsymbol \rho})\hat e_\sigma$
In Eq.~(\ref{Eq_HFS}), 
\begin{equation}
\xi_{\mathbf{k},q}^{\lambda,\sigma}=\sqrt{\frac{\omega_\sigma}{\omega_\mathrm{pt}(\mathbf k,q)}}P_{\lambda,\sigma}(\mathbf k,q)    
\end{equation}
corresponds to the part of the light-matter coupling which is linear on the photon vector potential  and
\begin{equation}
\Xi_{\mathbf{k},q,q^\prime}^{\lambda,\lambda'}=\sum_\sigma \xi_{\mathbf{k},q}^{\lambda,\sigma}\xi_{\mathbf{k},q}^{\lambda',\sigma},
\end{equation}
is the contribution to the light-matter coupling depending on $A^2$. The polarization function $P_{\lambda,\sigma}(\mathbf k,q)$ is given by
%\begin{equation}
%\label{eq:LightPolarizationMonolayer}
%\left\{\begin{matrix}
%P_{\lambda = s,\sigma}(\mathbf k,\mathbf q_z)=(\mathbf{\hat z}\times \mathbf{\hat k})\cdot \mathbf{\hat \sigma} \\
%P_{\lambda = p,\sigma}(\mathbf k,\mathbf q_z)=\frac{-q(\mathbf{\hat k} \cdot \mathbf{\hat \sigma}) +k(\mathbf{\hat z}\cdot\mathbf{\hat \sigma})}{
%\sqrt{k^2+q^2}}
%\end{matrix}\right.,
%\end{equation}
\begin{equation}
\label{eq:LightPolarizationMonolayer}
\left\{\begin{matrix}
P_{s,T}(\mathbf k,q)=&1 \\
P_{p,L}(\mathbf k,q)=&-\frac{q}{
\sqrt{k^2+q^2}}\\
P_{p,z}(\mathbf k,q)=&\frac{k}{
\sqrt{k^2+q^2}}
\end{matrix}\right.,
\end{equation}
where $\lambda=s$ correspond to photons polarized in the in-plane transverse direction whereas $\lambda=p$ corresponds to photons polarized along the $\hat{\mathbf k}\hat{\mathbf{z}}$ plane. Note that the transverse matter mode couples only to $s$ polarized photons while the longitudinal matter mode couples to $p$ photons, \textit{i. e.}, $P_{s,L}(\mathbf k,q)=P_{s,z}(\mathbf k,q)=
P_{p,T}(\mathbf k,q)=0$. 

%\section{Methods \label{sec:methods}}
\subsection{Electric-Field \label{ssec:Efield}}

The electric field operator is $\hat{\mathbf E}(\mathbf r,t)=-\nabla\hat\varphi(\mathbf r,t)-\frac{1}{c}\dot{\hat{\mathbf{A}}}(\mathbf r,t)$. In the presence of the 2D material, the dynamics of the fields are determined by the full polaritonic Hamiltonian. This requires that the electromagnetic potentials are written in terms of polaritonic operators. The total electric field is then  $\hat E_j(\mathbf r,t)=\sum_{\mathbf k,\tau}E_{j,\tau}(\mathbf k,\mathbf r)\hat\psi_{\tau,\mathbf k}(t)+h.c.$ with $E_{j,\tau}(\mathbf k,\mathbf r)=E^{sc}_{j,\tau}(\mathbf k,\mathbf r)+E^{pt}_{j,\tau}(\mathbf k,\mathbf r)$, where
\begin{equation}
E^{sc}_{j,\tau}(\mathbf k,\mathbf r)=(i^{\delta_{j,L}})(- s(z))^{\delta_{j,z}}\frac{g_{\mathrm{2D}}}{\omega_T} k\lambda_T {\sqrt\frac{\omega_T}{\omega_{\tau,\mathbf k}}} \mathrm  E_0(\omega_{\tau,\mathbf k}) 
\left(U^{-1}_{L,\tau}(\mathbf k)+V^{-1}_{L,\tau}(\mathbf k)\right)e^{i\mathbf k\cdot  \boldsymbol\rho-k|z|},
\end{equation}
is the contribution of the scalar potential, and
\begin{eqnarray}
E^{pt}_{j,\tau}(\mathbf k,\mathbf r)=i\mathrm E_0(\omega_{\tau,\mathbf k})\left[U^{-1}_{j,\tau}(\mathbf k,z)- V^{-1}_{j,\tau}(\mathbf k,z)\right] e^{i\mathbf k\cdot\boldsymbol \rho},
\end{eqnarray}
gives the vector potential (photonic) contribution to the electric field. The later is written in terms of 
\begin{eqnarray}
U^{-1}_{j,\tau }(\mathbf k,z)=\frac{1}{\sqrt{N_q}}\sum_{\lambda,q}\sqrt{\frac{\omega_{\tau,\mathbf k}}{\omega_{\mathbf k,q}}}U^{-1}_{\lambda,\tau}(\mathbf k,q) P_{\lambda,j}(\mathbf k,q)e^{iqz}\\
V^{-1}_{j,\tau }(\mathbf k,z)=\frac{1}{\sqrt{N_q}} \sum_{\lambda,q} \sqrt{\frac{\omega_{\tau,\mathbf k}}{\omega_{\mathbf k,q}}}V^{-1}_{\lambda,\tau}(\mathbf k,q) P_{\lambda,j}(\mathbf k,q)e^{iqz}.
\end{eqnarray}

The electric fields in this work are conveniently written in terms of a single-photon electric field
\begin{eqnarray}
\mathrm E_0(\omega_{\tau,\mathbf k})=\sqrt{\frac{\hbar\omega_{\tau,\mathbf k}}{2\epsilon_0 A\lambda_T}},    
\end{eqnarray}
normalized in the quantization volume $A\lambda_T$ and conveniently evaluated at the polariton frequency $\omega_{\tau,\mathbf k}$. 

\subsection{Longitudinal mode frequency renormalization}

\subsection{Evaluating $\Omega_\mathrm{2D}$}

We can use different strategies to try and evaluate $\Omega_\mathrm{2D}$ on different systems. If the 2D polaritonic dispersion is available in the literature, it is possible to estimate the coupling parameter by fitting the equation $\omega_L(k)=\sqrt{\omega_T^2+\Omega_\mathrm{2D}ck}$ to the obtained dispersion. In some cases, the effect of the surrounding dielectric function may have to be taken into account.\cite{Sohier2017} 

For systems on which polaritons have not been studied in the small thickness limit, an alternative strategy is to consider that the oscillator strength per material layer is independent of the layer thickness. In that case, the direct relation 
\begin{equation}
    \Omega_\mathrm{2D}=2\pi \frac{w}{\lambda_T}\frac{g_\mathrm{3D}^2}{\omega_T},
\end{equation}
can be obtained, where $g_\mathrm{3D}$ is the bulk light-matter coupling and $w$ is the width of a monolayer of the material.

\subsection{Width limit}
Within this simplified model, we considered the charge distribution to be infinitesimaly thin. This allows for any photon, regardless of its out-of-plane wavevector, to efficiently couple to the in-plane excitations. In a realistic 2D system the material can have a finite width. Furthermore, in many cases, the surface charge density can be approximately proportional to the width. The effective coupling of excitations in a 2D material with width $W$ to a photon with out-of-plane wavevector $q_z$ is
\begin{equation}
\Omega_\mathrm{eff}(W)=\Omega_\mathrm{2D}\frac{\sin(q_z W)}{q_zw}.   
\end{equation}
The results discussed here correspond to the limit of For $W=w<<2\pi/q_z$ for all relevant values of $q_z$. Since most relevant photons dressing the polaritonic states occur for $q_z\sim 2\pi/\lambda_T$, the proposed model should be applicable even for multilayered 2D materials, as long as the condition $W<<\lambda_T$ holds. This leads to the simple  

The approximation
\begin{equation}
\frac{\sin(2\pi W/\lambda_T)}{2\pi w} \lambda_T \approx \frac{W}{w} -\frac{4\pi^2}{6}\frac{W^3}{w\lambda_T^2}+\dots   
\end{equation}
to only the first order is within 95$\%$ precision for angles below $\pi/6$. Taking this as a limit to the accuracy of the model leads to 
\begin{equation}
W < \frac{\lambda_T }{12}     
\end{equation}

Together with the assumption that the charge densities per layer are constant, we can find the limiting effective coupling obtainable by stacking up layers of a given 2D material to the formation of 2D polaritons.
\begin{equation}
    \Omega^{max}_\mathrm{eff}= \frac{\pi}{6}\frac{g_\mathrm{3D}^2}{\omega_T},
\end{equation}
which corresponds approximately to $\frac{\Omega_\mathrm{eff}^{max}}{\omega_T}\approx 0.7 (\frac{g}{\omega_T})^2$. However, depending on the type of excitation, other factors may limit the maximum material width for which the proposed model should be applicable. For instance, in thicker samples, the optical response of the material in the out-of-plane direction deviates from unity and may play an important role. Further, the high-frequenccy dielectric function of the 2D material shorten the wavelength of the photon inside the 2D system, leading to further decoupling to the far-field photons.

%only occurs if the next term is weak, leading to the condition
%\begin{equation}
%\frac{2\pi^2}{3}\frac{W^2}{\lambda_T^2} << 1  
%\end{equation}
%adn finally to
%\begin{equation}
%W<< \sqrt{3/2} \frac{\lambda_T}{\pi}\approx 0.38\lambda_T     
%\end{equation}

%We have used these strategies to produce table

%\begin{table}[]
 %   \centering
 %   \begin{tabular}{c|ccccccc}
 %       Material & Excitation & $g_\mathbf{3D}$ (eV) & $\omega_T $ (eV) & width (nm) & $\Omega_\mathrm{2D}/\omega_T$ & ref.\\\hline
 %       hBN & phonon & 0.05 & 0.17 & $0.66$ & 5E-5 & [] \\  Squarilium & exciton &1.2 & 1.7 & 0.6 & 5E-3 &     
 %   \end{tabular}
 %   \caption{Caption}
 %   \label{tab:placeholder}
%\end{table}

%The longitudinal mode eigenfrequencies is renormalized due to the interaction with the longitudinal electric field (Coulomb interaction) created by the charge distribution of the mode. 
%The renormalized frequencies are by a Boguliobov transformation, see S.I., $\omega_L(k)=\sqrt{\omega_T^2+\Omega_\mathrm{2D}ck}$. This result leads to a simple way to estimate the effective coupling parameter experimentally as
%\begin{equation}
%g_\mathrm{2D}\approx\left.\sqrt{\frac{1}{\lambda_T} \frac{\partial \omega_L^2}{\partial k}}\right|_{k>>k_T}.
%\end{equation}

\subsection{Shrinking \label{ssec:shrink}}

The localization of the polaritonic electric fields in the in-plane direction can be evaluated by comparing the in-plane wavevector $k$ to the photon wavevector at polariton frequency $\omega_\sigma(k)/c$, from which we define the in-plane shrinking factor 
\begin{equation}
s_{\sigma}^{\parallel}(k)=ck/\omega_\sigma(k).    
\end{equation}
The out-of-plane shrinking factor, on the other hand, can be evaluated by comparing the attenuation (out-of-plane wavevector) to the free-photon wavevector, \begin{equation}
s_{\sigma}^\perp(k)=c\beta/\omega_\sigma(k).    
\end{equation}
From which we obtain the relation 
\begin{equation}
s_{\sigma}^{\parallel}(k)^2-s_{\perp}^{\parallel}(k)^2=1    
\end{equation}

\subsection{Squeezing\label{ssec:squeeze}}
The electric field operator can also be written as
\begin{eqnarray}
\hat E_{j}(\mathbf r,t)=\sum_{\mathbf k,\tau}\frac{|E_{j,\tau}(\mathbf k)|}{\sqrt 2}\left[\hat{\mathcal C}_{j,\tau}(\mathbf k)\cos(\omega_{\tau,\mathbf k} t)+\right.\nonumber \\\left.\hat{\mathcal S}_{j,\tau}(\mathbf k)\sin(\omega_{\tau,\mathbf k} t)\right]
\end{eqnarray}
where
\begin{eqnarray}
\hat{\mathcal C}_{j,\tau}(\mathbf k)=\frac{1}{\sqrt 2}\left(e^{i\theta}\hat\psi_{\mathbf k}+e^{-i\theta}\hat \psi_{\tau,-\mathbf k}+e^{-i\theta}\hat\psi^\dag_{\tau,\mathbf k}+e^{i\theta}\hat\psi _{\tau,-\mathbf k} \right)\nonumber  \\
\hat{\mathcal S}_{j,\tau}(\mathbf k)=\frac{i}{\sqrt 2}\left(e^{i\theta}\hat\psi_{\mathbf k}+e^{-i\theta}\tilde \psi_{\tau,-\mathbf k}-e^{-i\theta}\hat\psi^\dag_{\tau,\mathbf k}-e^{i\theta}\hat\psi_{\tau,-\mathbf k} \right),
\end{eqnarray}
where $\theta$ is short for $\theta_{j,\tau}(\mathbf k)$, the phase of the complex valued function $E_{j,\tau}(\mathbf k)=|E_{j,\tau}(\mathbf k)|e^{i\theta_{j,\tau}(\mathbf k)}$. The quadratures can also be written in terms of the bare excitation operators
\begin{eqnarray}
\hat{\mathcal C}_{j,\tau}(\mathbf k)=\sum_l\frac{e^{i\theta_{j,\tau}(\mathbf k)}}{\sqrt 2} [U_{\tau,l}(\mathbf k)+ V_{\tau,l}(\mathbf k)]\phi_{l,\mathbf k}\\ \nonumber +\frac{e^{-i\theta_{j,\tau}(\mathbf k)}}{\sqrt 2}[U^\ast_{\tau,l}(\mathbf k)+ V^\ast_{\tau,l}(\mathbf k)]\phi^\dag_{l,\mathbf k} \\ +\nonumber \frac{e^{-i\theta_{j,\tau}(\mathbf k)}}{\sqrt 2}[U^\ast_{\tau,l}(\mathbf k)+V^\ast_{\tau,l}(\mathbf k)]\phi_{l,-\mathbf k} \\\nonumber + \frac{e^{i\theta_{j,\tau}(\mathbf k)}}{\sqrt 2} [U_{\tau,l}(\mathbf k)+ V_{\tau,l}(\mathbf k)]\phi^\dag_{l,-\mathbf k}.
\end{eqnarray}
\begin{eqnarray}
\hat{\mathcal S}_{j,\tau}(\mathbf k)=\sum_l\frac{ie^{i\theta_{j,\tau}(\mathbf k)}}{\sqrt 2} [U_{\tau,l}(\mathbf k)- V_{\tau,l}(\mathbf k)]\phi_{l,\mathbf k}\\ \nonumber -\frac{ie^{-i\theta_{j,\tau}(\mathbf k)}}{\sqrt 2}[U^\ast_{\tau,l}(\mathbf k)- V^\ast_{\tau,l}(\mathbf k)]\phi^\dag_{l,\mathbf k} \\ +\nonumber \frac{ie^{-i\theta_{j,\tau}(\mathbf k)}}{\sqrt 2}[U^\ast_{\tau,l}(\mathbf k)- V^\ast_{\tau,l}(\mathbf k)]\phi_{l,-\mathbf k} \\\nonumber - \frac{ie^{i\theta_{j,\tau}(\mathbf k)}}{\sqrt 2} [U_{\tau,l}(\mathbf k)- V_{\tau,l}(\mathbf k)]\phi^\dag_{l,-\mathbf k}.
\end{eqnarray}
here $l$ runs for all bare excitations of the system \textit{i.e.} $\phi_{l,\mathbf k}\equiv\phi_{l,\mathbf k}(0)=b_{\mathbf k,z},~b_{\mathbf k,L},~c_{\mathbf k,q,s},~c_{\mathbf k,q,p},...$. For a bare photon coherent state $|\alpha_{\mathbf k,q,\lambda}\rangle$, the fluctuations of the quadrature operators are 
\begin{eqnarray}
\Delta X_{\tau}=\sqrt{\sum_l\left|\left(U_{\tau,l}(\mathbf k)+ V_{\tau,l}(\mathbf k)\right)\right|^2}\\
\Delta Y_{\tau}=\sqrt{\sum_l\left|\left(U_{\tau,l}(\mathbf k)- V_{\tau,l}(\mathbf k)\right)\right|^2}, 
\end{eqnarray}
following the uncertainty relation $\Delta X_{\tau}\Delta Y_{\tau}\ge 1$. The squeezing factor $r_{\tau}(\mathbf k)=\log\frac{\Delta X_{j,\tau}}{\Delta Y_{j,\tau}}$ is
\begin{equation}
    r_{\tau}(\mathbf k)= \frac{1}{2}\tanh^{-1} \frac{\mathrm{Re}[2U_{\tau,s}(\mathbf k)V_{\tau,s}(\mathbf k)]}{1+2\sum_s |V_{\tau,s}(\mathbf k)|^2}.
\end{equation}

Finally, in the ground state, the expectation value of both quadrature operators vanish and the expected electric field distribution is of a Gaussian form centered at $X=Y=0$ and with widths given by $\Delta X$ and $\Delta Y$.

\bibliography{mendeley2Dpol}

%\bibliography{sn-bibliography}
\section{Extended Data}

%\begin{figure*}
5    \centering
  %  \includegraphics[width=4.2cm]{2D_polariton_model.png}
 %   \begin{subfigure}
 %   \includegraphics[width=8cm]{Fig_Squeeze_g.pdf}     
 
 %\end{subfigure}
    
 %  \includegraphics[width=9cm]{Fig_Emap_hop.pdf}
%    \caption{Squeeze factor of the TE (pink) and TM (green) polariton modes as a function of the reduced coupling parameter $g/\omega_T$ for two values of $k$. }
%    \label{fig:dispersion}
%\end{figure*}

\begin{figure*}[h]
    \centering
  %  \includegraphics[width=4.2cm]{2D_polariton_model.png}
 %   \begin{subfigure}
    \includegraphics[width=8cm]{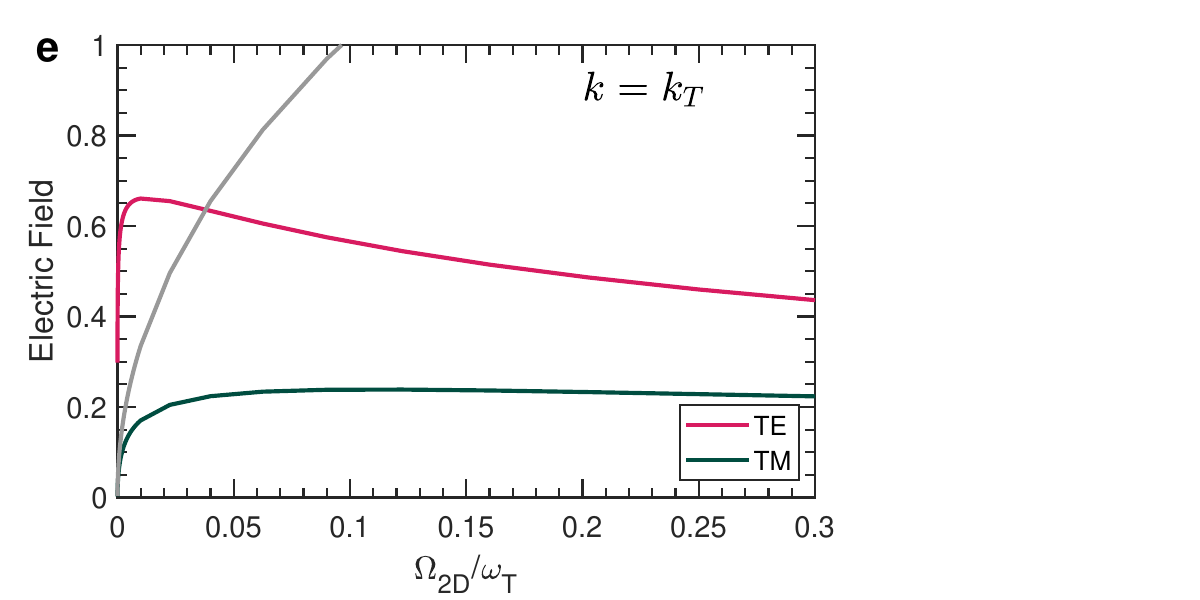}     
 
 %\end{subfigure}
    
 %  \includegraphics[width=9cm]{Fig_Emap_hop.pdf}
    \caption{Frequency of the TE (pink) and TM (green) polariton modes as a function of the reduced coupling parameter $g/\omega_T$ for two values of $k$. }
    \label{fig:frequency_g}
\end{figure*}

\end{document}